\documentclass[pra,superscriptaddress,twocolumn,showpacs,preprintnumbers,amsmath,amssymb,longbibliography]{revtex4} 

\usepackage[dvipdfmx]{graphicx}
\usepackage{dcolumn,upgreek}
\usepackage{bm}
\usepackage{color}

\def\degree{\mbox{$^\circ$}}

\def\ii{{\mathrm{i}}}

\def\HH{{\mathrm{H}}}
\def\VV{{\mathrm{V}}}

\def\ket#1{|#1\rangle}

\def\bracketi#1#2{\langle #1 | #2 \rangle}

\begin{document}


\title{Observation of nonlinear variations in three-vertex geometric phase\\ in two-photon polarization qutrit}

\author{Kazuhisa Ogawa}
\email{ogawa@giga.kuee.kyoto-u.ac.jp}
\affiliation{%
Department of Electronic Science and Engineering, Kyoto University, Kyoto 615-8510, Japan
}%
 
\author{Shuhei Tamate}
\affiliation{%
National Institute of Informatics, Hitotsubashi, Chiyoda-ku, Tokyo 101-8430, Japan
}%
\author{Hirokazu Kobayashi}
\affiliation{%
Department of Electronic and Photonic System Engineering,\\
Kochi University of Technology, Tosayamada-cho, Kochi 782-8502, Japan
}%
\author{Toshihiro Nakanishi}
\affiliation{%
Department of Electronic Science and Engineering, Kyoto University, Kyoto 615-8510, Japan
}%
\author{Masao Kitano}
\affiliation{%
Department of Electronic Science and Engineering, Kyoto University, Kyoto 615-8510, Japan
}%

\date{\today}

\begin{abstract}

We experimentally observed nonlinear variations in the three-vertex geometric phase in a two-photon polarization qutrit.
The three-vertex geometric phase is defined by three quantum states, which generally forms a three-state (qutrit) system.
By changing one of the three constituent states, we observed two rapid increases in the three-vertex geometric phase.
The observed variations are inherent in a three-state system and cannot be observed in a two-state system.
We used a time-reversed two-photon interferometer to measure the geometric phase with much more intense signals than those of a typical two-photon interferometer. 

\end{abstract}

\maketitle

\section{Introduction}

The geometric phase is a fundamental concept in many areas of physics.
It was discovered by Berry~\cite{berry1984quantal} as an additional phase factor that emerges in adiabatic and cyclic evolution of a quantum state.
The definition of the geometric phase was extended to the non-adiabatic~\cite{aharonov1987phase} and non-cyclic~\cite{samuel1988general} cases and was finally generalized on the basis of kinematic ideas by Mukunda and Simon~\cite{mukunda1993quantum}.
In their formulation, the geometric phase is defined by a trajectory on the quantum state space and is represented as a sum of the following \textit{three-vertex geometric phases}:
\begin{align}
\gamma(\psi_1,\psi_2,\psi_3):=\arg\bracketi{\psi_1}{\psi_3}\bracketi{\psi_3}{\psi_2}\bracketi{\psi_2}{\psi_1},
\end{align}
which is defined by three quantum states~\cite{pancharatnam1956generalized}.
Therefore, the three-vertex geometric phase is regarded as a fundamental building block of an arbitrary geometric phase.

The three-vertex geometric phase is ubiquitous in various physical systems involving three different states.
In optical systems, the three-vertex geometric phase appears in an additional phase factor after three polarization projections~\cite{pancharatnam1956generalized} or three reflections~\cite{kitano1987comment}, and in the interference patterns of three differently polarized beams~\cite{kobayashi2010direct}.
In the problem distinguishing three quantum states, the three-vertex geometric phase is an important factor characterizing their distinguishability~\cite{PhysRevA.62.012301,PhysRevLett.108.250502,PhysRevA.82.032338}.
In addition, the quantum eraser~\cite{PhysRevA.25.2208} and weak-value amplification~\cite{PhysRevLett.60.1351} are related to the three-vertex geometric phase defined by the initial, intermediate, and final states in the systems~\cite{JPSJ.80.034401,tamate2009geometrical}.

The three-vertex geometric phase has been widely studied in a two-state (qubit) system.
In a two-state system, the three-vertex geometric phase is geometrically represented as the area of a spherical triangle formed by the three constituent states on the Bloch (Poincar\'{e}) sphere~\cite{pancharatnam1956generalized}. 
Various nonlinear behaviors of the three-vertex geometric phase in a two-state system have been investigated using the Bloch sphere representation and observed in several optical experiments~\cite{PhysRevLett.71.1530,tewari1995four,li1999experimental,PhysRevE.60.2322,JPSJ.80.034401}.

However, the three arbitrary states that define a three-vertex geometric phase generally span a three-dimensional Hilbert space; therefore, we need to treat a three-state (qutrit) system to investigate the general properties of the three-vertex geometric phase.
In our previous study~\cite{tamate2011bloch}, we constructed a geometric representation of the three-vertex geometric phase in a three-state system on the Bloch sphere.
Using the Bloch sphere representation, we predicted some nonlinear variations in the three-vertex geometric phase inherent in a three-state system.

In this paper, we experimentally observe the nonlinear variations in the three-vertex geometric phase inherent in a three-state system with an optical interferometer.
We employ the polarizations of two photons in the same spatiotemporal mode (a two-photon polarization qutrit) as a three-state system~\cite{Physrevlett.93.230503,PhysRevLett.100.060504,PhysRevA.71.062337,PhysRevA.76.012319}.
In our setup, the three-vertex geometric phase exhibits two rapid increases with respect to a change in one of the three constituent states.
We use a time-reversed two-photon interferometer~\cite{PhysRevA.88.063813} for the measurement.
Unlike the typical method of measuring the geometric phase in two-photon polarization~\cite{kobayashi2011nonlinear,PhysRevA.52.2551,PhysRevLett.112.143603}, our setup can obtain vastly more intense signals and can be implemented using classical light.

This paper is organized as follows. 
In Sec.~\ref{sec:theory}, we briefly review the theory of the three-vertex geometric phase in a three-state system investigated in Ref.~\cite{tamate2011bloch}.
In Sec.~\ref{sec:experimental-setup}, we describe the experimental observation of the nonlinear variations in the three-vertex geometric phase. 
We also discuss the advantages of using a time-reversed two-photon interferometer for the experiments.
Finally, we summarize the findings of our study in Sec.~\ref{sec:conclusion}.

\section{Theory}\label{sec:theory}

We describe the Bloch sphere representation of the three-vertex geometric phase in a three-state system~\cite{tamate2011bloch}.
We also derive the nonlinear variations in the three-vertex geometric phase inherent in a three-state system, which are experimentally observed in Sec.~\ref{sec:experimental-setup}.

\begin{figure}
\centering
\includegraphics[width=8cm]{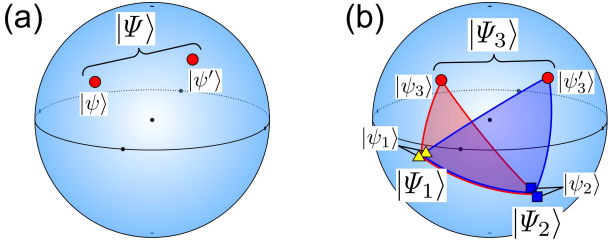}
\caption{(Color online) (a) Bloch sphere representation of the symmetrized two-qubit state $\ket{\varPsi}$. (b) Bloch sphere representation of the three-vertex geometric phase of the standard triplet $\gamma(\varPsi_1,\varPsi_2,\varPsi_3)$, which is proportional to the sum of the area of the two spherical triangles.}\label{fig:1}
\end{figure}

A three-state system can be identified in terms of a symmetrized two-qubit system. 
The symmetrized two-qubit state $\ket{\varPsi}$ is described as 
\begin{align}
\ket{\varPsi}=k(\ket{\psi}\ket{\psi'}+\ket{\psi'}\ket{\psi}),
\end{align}
where $\ket{\psi}$ and $\ket{\psi'}$ are qubit states, and $k$ is a normalization factor (in what follows, we omit $k$ for simplicity).
$\ket{\varPsi}$ can be uniquely depicted as the two points corresponding to $\ket{\psi}$ and $\ket{\psi'}$ on the Bloch sphere (Majorana's stellar representation \cite{majorana,PhysRevLett.113.240403,PhysRevLett.108.240402,hannay1998berry}), as shown in Fig.~\ref{fig:1}(a).

To visualize the three-vertex geometric phase on the Bloch sphere, we consider the following \textit{standard triplet}:
\begin{align}
\begin{split}
\ket{\varPsi_1}&=\ket{\psi_1}\ket{\psi_1},\quad
\ket{\varPsi_2}=\ket{\psi_2}\ket{\psi_2},\\
&\ket{\varPsi_3}=\ket{\psi_3}\ket{\psi'_3}+\ket{\psi'_3}\ket{\psi_3},\label{eq:4}
\end{split}
\end{align} 
where $\ket{\varPsi_1}$ and $\ket{\varPsi_2}$ are product states, and $\ket{\varPsi_3}$ is an arbitrary symmetrized two-qubit state.
Although the standard triplet is a special set of three states, any set of three states can be mapped onto a standard triplet by applying the proper unitary transformation \cite{tamate2011bloch}.
The three-vertex geometric phase of the standard triplet is expressed as the sum of two three-vertex geometric phases in two-state systems:
\begin{align}
\gamma(\varPsi_1,\varPsi_2,\varPsi_3)&=\gamma(\psi_1,\psi_2,\psi_3)+\gamma(\psi_1,\psi_2,\psi'_3).\label{eq:2}
\end{align}
Because a three-vertex geometric phase in a two-state system is equal to $-1/2$ times the area of a spherical triangle on the Bloch sphere \cite{pancharatnam1956generalized}, the three-vertex geometric phase of the standard triplet can be depicted as the area of two spherical triangles on the Bloch sphere, as shown in Fig.~\ref{fig:1}(b). 
In this manner, we can represent an arbitrary three-vertex geometric phase in a three-state system on the Bloch sphere.  
 
\begin{figure}
\centering
\includegraphics[width=8.5cm]{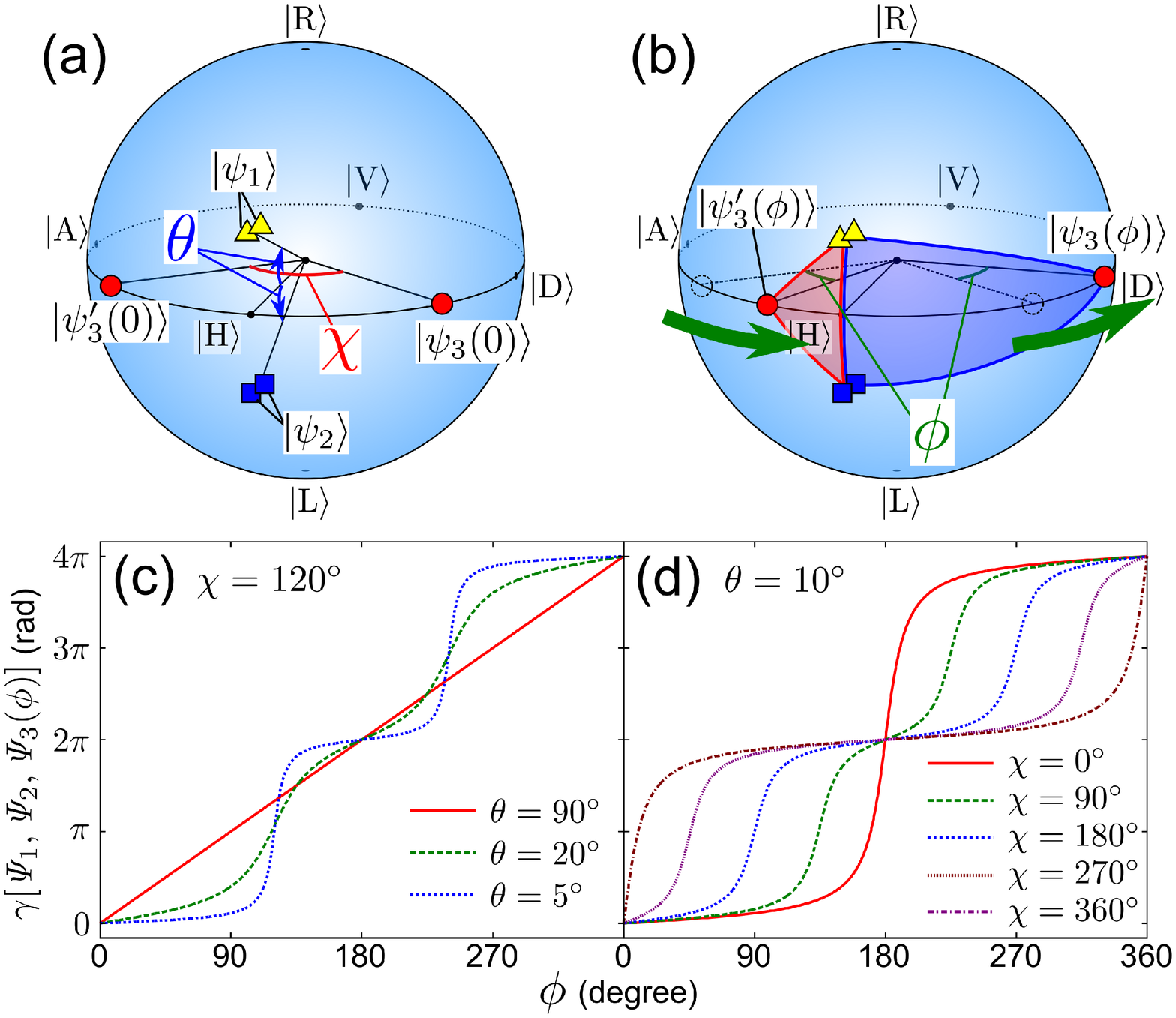}
\caption{(Color online) (a) Bloch sphere representation of the states given by Eqs.~(\ref{eq:3})--(\ref{eq:13}) when $\phi=0\degree$. $\theta$ is the half-angle between the states $\ket{\psi_1}$ and $\ket{\psi_2}$, and $\chi$ is the angle between the states $\ket{\psi_3(\phi)}$ and $\ket{\psi_3'(\phi)}$. 
(b) Bloch sphere representation of the geometric phase $\gamma[\varPsi_1,\varPsi_2,\varPsi_3(\phi)]$ when the two red circles $\ket{\psi_3(\phi)}$ and $\ket{\psi_3'(\phi)}$ are rotated along the equator. 
When the red circles pass through the reverse side of the Bloch sphere, the area of the spherical triangles increases rapidly. 
In addition, as the angle between the yellow triangles and the blue squares decreases, the area of the two spherical triangles increases more rapidly. 
(c), (d) The variations in $\gamma[\varPsi_1,\varPsi_2,\varPsi_3(\phi)]$ with respect to $\phi$, for several values of $\theta$ when $\chi=120\degree$ (c) and for several values of $\chi$ when $\theta=10\degree$ (d). 
}\label{fig:2}
\end{figure}

We next derive the nonlinear variations in the three-vertex geometric phase inherent in a three-state system from the Bloch sphere representation. 
We employ a two-photon polarization qutrit as a three-state system.
$\ket{\HH}$ and $\ket{\VV}$ denote the horizontal and vertical polarization states, respectively. 

We now consider the following standard triplet of two-photon polarization qutrits:
\begin{align}
\ket{\varPsi_1}&=\ket{\psi_1}\ket{\psi_1},\quad
\ket{\varPsi_2}=\ket{\psi_2}\ket{\psi_2},\label{eq:14}\\
\ket{\varPsi_3(\phi)}&=\ket{\psi_3(\phi)}\ket{\psi'_3(\phi)}+\ket{\psi'_3(\phi)}\ket{\psi_3(\phi)},\label{eq:15}
\end{align}
where
\begin{align}
\ket{\psi_1}&:=\cos(\theta/2)\ket{\HH}+\ii\sin(\theta/2)\ket{\VV},\label{eq:3}\\
\ket{\psi_2}&:=\cos(\theta/2)\ket{\HH}-\ii\sin(\theta/2)\ket{\VV},\label{eq:9}\\
\ket{\psi_3(\phi)}&:=\cos\left(\frac{\chi}{4}+\frac{\phi}{2}\right)\ket{\HH}+\sin\left(\frac{\chi}{4}+\frac{\phi}{2}\right)\ket{\VV},\label{eq:12}\\
\ket{\psi'_3(\phi)}&:=\cos\left(\frac{\chi}{4}-\frac{\phi}{2}\right)\ket{\HH}-\sin\left(\frac{\chi}{4}-\frac{\phi}{2}\right)\ket{\VV}.\label{eq:13}
\end{align} 
This standard triplet is depicted in Fig.~\ref{fig:2}(a). 
The parameters $\theta$ and $\chi$ are fixed at certain values.
We change $\phi$ to rotate the two red circles $\ket{\psi_3(\phi)}$ and $\ket{\psi'_3(\phi)}$ along the equator on the Bloch sphere, as shown in Fig.~\ref{fig:2}(b).
The three-vertex geometric phase $\gamma\left[\varPsi_1,\varPsi_2,\varPsi_3(\phi)\right]$ is calculated as
\begin{align}
\gamma\left[\varPsi_1,\varPsi_2,\varPsi_3(\phi)\right]
&=\gamma\left[\psi_1,\psi_2,\psi_3(\phi)\right]+\gamma\left[\psi_1,\psi_2,\psi_3'(\phi)\right],\label{eq:5}\\
\gamma\left[\psi_1,\psi_2,\psi_3(\phi)\right]
&=-2\tan^{-1}\left[\tan\frac{\theta}{2}\cdot\tan\left(\frac{\chi}{4}+\frac{\phi}{2}\right)\right],\label{eq:7}\\
\gamma\left[\psi_1,\psi_2,\psi_3'(\phi)\right]
&=2\tan^{-1}\left[\tan\frac{\theta}{2}\cdot\tan\left(\frac{\chi}{4}-\frac{\phi}{2}\right)\right].\label{eq:8}
\end{align}

The variations in $\gamma[\varPsi_1,\varPsi_2,\varPsi_3(\phi)]$ with respect to $\phi$ for several values of $\theta$ and $\chi$ are shown in Fig.~\ref{fig:2}(c) and (d). 
These figures indicate that the variations in $\gamma[\varPsi_1,\varPsi_2,\varPsi_3(\phi)]$ exhibit two rapid increases by $2\pi$ at the angles $\phi=180\degree\pm\chi/2$, and as the angle $\theta$ decreases, the geometric phase increases more rapidly. 
These rapid variations in $\gamma[\varPsi_1,\varPsi_2,\varPsi_3(\phi)]$ are interpreted as nonlinear variations in the area of the two spherical triangles on the Bloch sphere.

\section{Experiments}\label{sec:experimental-setup}

We next describe our experimental observation of the nonlinear variations in the three-vertex geometric phase in a three-state system derived in Sec.~\ref{sec:theory}.
In Sec.~\ref{sec:experimental-setup-1}, we describe our experimental setup for measuring the geometric phase using an optical interferometer.
In Sec.~\ref{sec:results}, we show the measured nonlinear variations in the geometric phase.
In Sec.~\ref{sec:summary-discussion}, we discuss the advantages of our experimental setup for measuring the geometric phase in two-photon polarization.

\subsection{Experimental setup}\label{sec:experimental-setup-1}

\begin{figure}
\centering
\includegraphics[width=8.5cm]{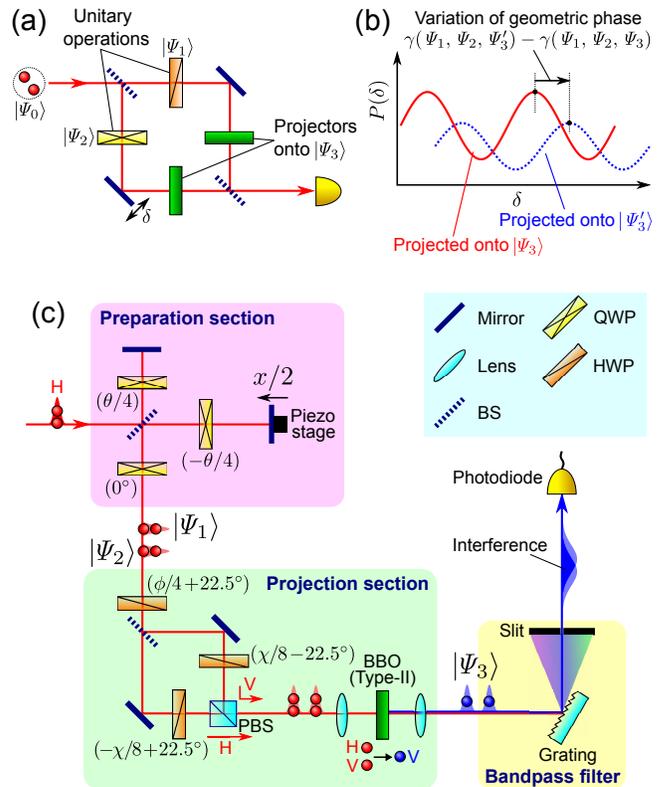}
\caption{(Color online) 
(a) Schematic illustration of setup for measuring the three-vertex geometric phase in a quantum eraser. 
(b) Projection probability $P(\delta)$ for different final internal states $\ket{\varPsi_3}$ and $\ket{\varPsi_3'}$. 
From the phase shift of the interference fringes, we can measure the variation in the three-vertex geometric phase $\gamma(\varPsi_1,\varPsi_2,\varPsi'_3)-\gamma(\varPsi_1,\varPsi_2,\varPsi_3)$.
(c) Experimental setup for measuring the three-vertex geometric phase in two-photon polarization qutrit. 
QWP, quarter-wave plate; HWP, half-wave plate; BS, (non-polarizing) beam splitter; PBS, polarizing beam splitter; BBO, $\upbeta$-barium borate crystal.
The values in the parentheses next to the QWPs and HWPs denote the angles of their fast axes from the horizontal axis. 
The parameter $\theta$ is adjusted by changing the angles of the QWPs.
The parameters $\chi$ and $\phi$ are adjusted by changing the angles of the HWPs.}
\label{fig:exp_1}
\end{figure}

In the experiment, we measure the three-vertex geometric phase using a quantum eraser~\cite{JPSJ.80.034401,kobayashi2011nonlinear}.
Let us consider the interferometer of a photon pair shown in Fig.~\ref{fig:exp_1}(a). 
The input photon pair with the initial two-photon polarization qutrit state $\ket{\varPsi_0}$ is first split into two arms by a beam splitter. 
The two-photon polarization qutrit states of the upper and lower mode are transformed into $\ket{\varPsi_1}$ and $\ket{\varPsi_2}$, respectively, by unitary operations.
Subsequently, both of the two-photon polarization qutrit states are projected onto $\ket{\varPsi_3}$, and the two path modes are combined by another beam splitter.
By changing the relative phase $\delta$ between the two path modes, we can observe interference fringes. 
When the final state $\ket{\varPsi_3}$ varies, the variation in the three-vertex geometric phase $\gamma(\varPsi_1,\varPsi_2,\varPsi_3)$ can be measured from a phase shift of the interference fringes, as shown in Fig.~\ref{fig:exp_1}(b). 

The actual experimental setup is shown in Fig.~\ref{fig:exp_1}(c), which implements the measurement method using a quantum eraser.

We used a femtosecond fiber laser (center wavelength 782\,nm, pulse duration 74.5\,fs, average power 54\,mW, repetition rate 100\,MHz) to create transform-limited pulsed light with horizontal polarization.
The input pulse enters the preparation section, which forms an unbalanced Michelson interferometer including three quarter-wave plates (QWPs).
The optical path difference $x$ between the two arms of the interferometer can be changed by a piezoelectric actuator and is adjusted to about 100\,$\upmu$m.
The two output pulses of the interferometer are substantially separated in time and do not interfere with each other.
After passing through the third QWP, the two-photon polarization qutrit states of the later and earlier pulses are transformed into $\ket{\varPsi_1}=\ket{\psi_1}\ket{\psi_1}$ and $\ket{\varPsi_2}=\ket{\psi_2}\ket{\psi_2}$ in Eq.~(\ref{eq:14}), respectively. 

The pulses next pass through the projection section, which consists of three half-wave plates (HWPs), polarizing and non-polarizing beam splitters (PBS and BS), and a 1-mm-long $\upbeta$-barium borate (BBO) crystal for collinear type-II sum-frequency generation (SFG).
The HWPs, PBS, and BS convert the polarizations $\ket{\psi_3(\phi)}$ and $\ket{\psi'_3(\phi)}$ in Eqs.~(\ref{eq:12}) and (\ref{eq:13}) into $\ket{\HH}$ and $\ket{\VV}$, respectively.
Subsequently, the BBO crystal converts only two photons with the two-photon polarization qutrit state $\ket{\HH}\ket{\VV}+\ket{\VV}\ket{\HH}$ into a sum-frequency photon. 
Therefore, the entire section projects the two-photon polarization qutrit state onto $\ket{\varPsi_3(\phi)}$ in Eq.~(\ref{eq:15}).

The two sum-frequency pulses are filtered to pass a 0.23-nm bandwidth centered around 391\,nm by a 1,200-lines/mm aluminum-coated diffraction grating followed by a slit. 
After the bandpass filter, the coherence lengths of the pulses are broadened, and the pulses interfere with each other. 
The optical power is measured by a Si photodiode (New Focus, Model 2151).

We measured the interference fringes as a function of the optical path difference $x$ for various values of $\theta$, $\chi$, and $\phi$, and derived the variations in the three-vertex geometric phase from the shifts of the fringes.

\subsection{Results}\label{sec:results}

\begin{figure}
\centering
\includegraphics[width=8.5cm]{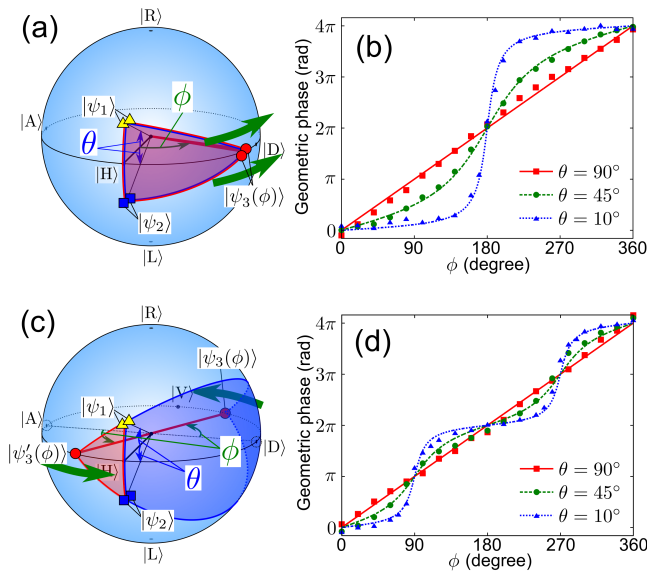}
\caption{
(Color online) Variations in the three-vertex geometric phase for $\chi=0\degree$ [(a), (b)] and $\chi=180\degree$ [(c), (d)].
(a), (c) Bloch sphere representation of the three-vertex geometric phase.
(b), (d) Measured variations in the geometric phase with respect to $\phi$.
As the Bloch sphere representation predicts, the geometric phase increases rapidly by $4\pi$ at $\phi=180\degree$ in (b) and by $2\pi$ at $\phi=90\degree$ and $270\degree$ in (d).
We can see that as the angle $\theta$ decreases, the geometric phase increases more rapidly.
}
\label{fig:graph_GP}
\end{figure}

We first measured the variations in the three-vertex geometric phase with respect to $\phi$ for several values of $\theta$ when $\chi=0\degree$ [Figs.~\ref{fig:graph_GP}(a), (b)] and $\chi=180\degree$ [Figs.~\ref{fig:graph_GP}(c), (d)].
From the Bloch sphere representation [Figs.~\ref{fig:graph_GP}(a), (c)], we can predict that the geometric phase increases rapidly by $4\pi$ at $\phi=180\degree$ when $\chi=0\degree$ and by $2\pi$ at $\phi=90\degree$ and $270\degree$ when $\chi=180\degree$.
Figures~\ref{fig:graph_GP}(b) and (d) show the measurement results, where the dots and lines denote the measurement data and the theoretical lines, respectively.
Because the measured geometric phase is a relative value with respect to a certain offset value, we determined the offset value by fitting the measurement data to the theoretical lines for every setting of the parameters $\theta$ and $\chi$.
The measurement results agree well with the theoretical prediction. 
As the angle $\theta$ decreases, the geometric phase increases more rapidly.

\begin{figure*}
\centering
\includegraphics[width=18cm]{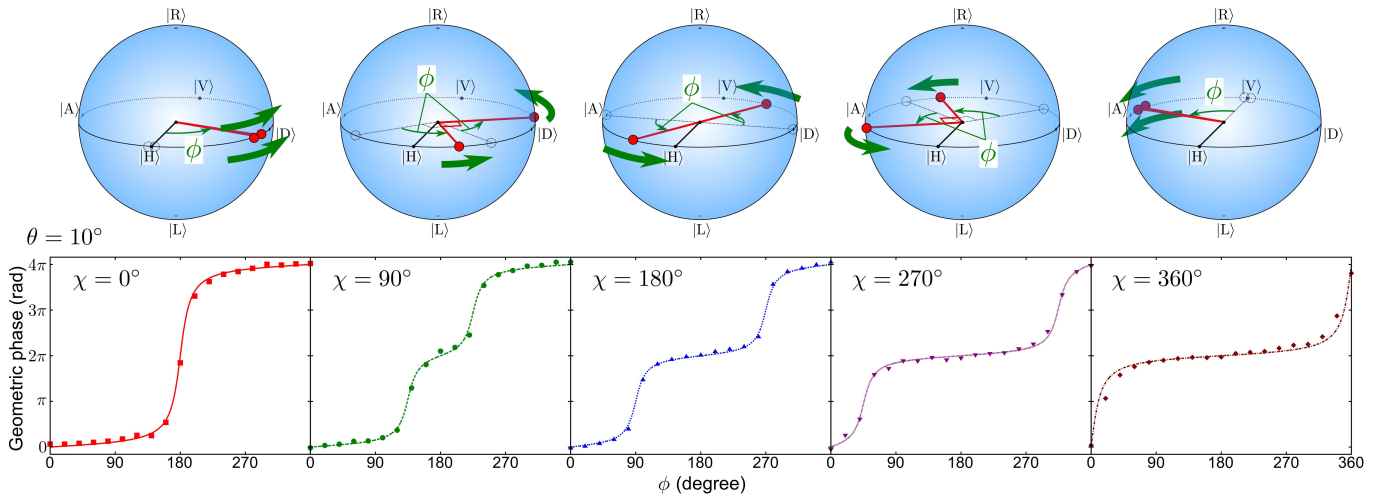}
\caption{(Color online) Variations in the three-vertex geometric phase with respect to $\phi$ for several values of $\chi$ when $\theta=10\degree$.
The upper panels are the Bloch sphere representations of $\ket{\psi_3(\phi)}$ and $\ket{\psi'_3(\phi)}$.
The graphs in the lower panels show the measurement results corresponding to each Bloch sphere.
As the Bloch sphere representation predicts, the locations of the two jumping points of the geometric phase depend on $\chi$.}
\label{fig:graph_GP_shift}
\end{figure*} 

We also measured the variations in the three-vertex geometric phase with respect to $\phi$ for several values of $\chi$ when $\theta=10\degree$.
From the Bloch sphere representation (the upper panels in Fig.~\ref{fig:graph_GP_shift}), we can predict that the two values of $\phi$ at which the geometric phase increases rapidly depend on $\chi$: $\phi=180\degree\pm\chi/2$.
The graphs in the lower panels in Fig.~\ref{fig:graph_GP_shift} show the measurement results, where the dots and lines denote the measurement data and the theoretical lines, respectively, and we determined the offset value of the measurement data in the same manner as described above.
We can see that the measurement results agree well with the theoretical prediction by our Bloch sphere representation, and the locations of the two jumping points depend on $\chi$.

\subsection{Discussion}\label{sec:summary-discussion}

Here we discuss the advantages of our experimental setup for measuring the geometric phase in two-photon polarization.

Such measurements have typically been made using a two-photon interferometer~\cite{kobayashi2011nonlinear,PhysRevA.52.2551,PhysRevLett.112.143603}.
In the two-photon interferometer, we need to generate photon pairs by spontaneous parametric down-conversion (SPDC) and to detect them by coincidence counting.
Because of the low efficiency of generation and detection of entangled photon pairs, the two-photon interferometer often suffers from weak output signals, which makes our estimation of the geometric phase uncertain.

In contrast, we employ a time-reversed two-photon interferometer \cite{PhysRevA.88.063813} for measuring the geometric phase. 
The time-reversed two-photon interferometer up-converts two photons into a sum-frequency photon by SFG instead of generating photon pairs by SPDC.
Because the output power of SFG is proportional to the square of the input power, we can observe vastly more intense interference signals.
Moreover, coincidence counting of photon pairs is not needed in the time-reversed two-photon interferometer; therefore, we can measure the geometric phase in two-photon polarization using a simpler setup.

In our experiment, the output signal power is minimized when $\theta=10\degree$, $\chi=0\degree$, and $\phi=180\degree$. 
Even in this condition, the measured average power of the interference fringes is 1.1\,pW, which corresponds to $2.1\times10^6$\,photons/s.
This output power is three orders of magnitude greater than that in previous experiments~\cite{kobayashi2011nonlinear,PhysRevA.52.2551,PhysRevLett.112.143603}.

\section{Conclusion}\label{sec:conclusion}

We presented optical experiments for measuring the three-vertex geometric phase in a two-photon polarization qutrit.
We experimentally demonstrated the nonlinear variations in the three-vertex geometric phase, which is inherent in a three-state system such as a two-photon polarization qutrit.  
The nonlinear variations are predicted by the Bloch sphere representation, and all the measurement results agree well with the theoretical prediction.
We also noted that our measurement method for the geometric phase using a time-reversed two-photon interferometer enables us to obtain vastly more intense output signals.
This measurement method can be used for high-intensity measurement of other properties of the geometric phase.  
We expect that this study will motivate the investigation of the geometric phase in high-dimensional systems and will enable new quantum optical technology using the geometric phase.

\section*{ACKNOWLEDGMENTS}

This research was supported by Japan Society for the Promotion of Science KAKENHI Grants No. 22109004 and No. 25287101.

\appendix

\newpage 


\begin{thebibliography}{31}
\expandafter\ifx\csname natexlab\endcsname\relax\def\natexlab#1{#1}\fi
\expandafter\ifx\csname bibnamefont\endcsname\relax
  \def\bibnamefont#1{#1}\fi
\expandafter\ifx\csname bibfnamefont\endcsname\relax
  \def\bibfnamefont#1{#1}\fi
\expandafter\ifx\csname citenamefont\endcsname\relax
  \def\citenamefont#1{#1}\fi
\expandafter\ifx\csname url\endcsname\relax
  \def\url#1{\texttt{#1}}\fi
\expandafter\ifx\csname urlprefix\endcsname\relax\def\urlprefix{URL }\fi
\providecommand{\bibinfo}[2]{#2}
\providecommand{\eprint}[2][]{\url{#2}}

\bibitem[{\citenamefont{Berry}(1984)}]{berry1984quantal}
\bibinfo{author}{\bibfnamefont{M.~V.} \bibnamefont{Berry}},
  \bibinfo{journal}{Proc. R. Soc. London A} \textbf{\bibinfo{volume}{392}},
  \bibinfo{pages}{45} (\bibinfo{year}{1984}).

\bibitem[{\citenamefont{Aharonov and Anandan}(1987)}]{aharonov1987phase}
\bibinfo{author}{\bibfnamefont{Y.}~\bibnamefont{Aharonov}} \bibnamefont{and}
  \bibinfo{author}{\bibfnamefont{J.}~\bibnamefont{Anandan}},
  \bibinfo{journal}{Phys. Rev. Lett.} \textbf{\bibinfo{volume}{58}},
  \bibinfo{pages}{1593} (\bibinfo{year}{1987}).

\bibitem[{\citenamefont{Samuel and Bhandari}(1988)}]{samuel1988general}
\bibinfo{author}{\bibfnamefont{J.}~\bibnamefont{Samuel}} \bibnamefont{and}
  \bibinfo{author}{\bibfnamefont{R.}~\bibnamefont{Bhandari}},
  \bibinfo{journal}{Phys. Rev. Lett.} \textbf{\bibinfo{volume}{60}},
  \bibinfo{pages}{2339} (\bibinfo{year}{1988}).

\bibitem[{\citenamefont{Mukunda and Simon}(1993)}]{mukunda1993quantum}
\bibinfo{author}{\bibfnamefont{N.}~\bibnamefont{Mukunda}} \bibnamefont{and}
  \bibinfo{author}{\bibfnamefont{R.}~\bibnamefont{Simon}},
  \bibinfo{journal}{Ann. Phys. (NY)} \textbf{\bibinfo{volume}{228}},
  \bibinfo{pages}{205} (\bibinfo{year}{1993}).

\bibitem[{\citenamefont{Pancharatnam}(1956)}]{pancharatnam1956generalized}
\bibinfo{author}{\bibfnamefont{S.}~\bibnamefont{Pancharatnam}},
  \bibinfo{journal}{Proc. Indian Acad. Sci. A} \textbf{\bibinfo{volume}{44}},
  \bibinfo{pages}{247} (\bibinfo{year}{1956}).

\bibitem[{\citenamefont{Kitano et~al.}(1987)\citenamefont{Kitano, Yabuzaki, and
  Ogawa}}]{kitano1987comment}
\bibinfo{author}{\bibfnamefont{M.}~\bibnamefont{Kitano}},
  \bibinfo{author}{\bibfnamefont{T.}~\bibnamefont{Yabuzaki}}, \bibnamefont{and}
  \bibinfo{author}{\bibfnamefont{T.}~\bibnamefont{Ogawa}},
  \bibinfo{journal}{Phys. Rev. Lett.} \textbf{\bibinfo{volume}{58}},
  \bibinfo{pages}{523} (\bibinfo{year}{1987}).

\bibitem[{\citenamefont{Kobayashi et~al.}(2010)\citenamefont{Kobayashi, Tamate,
  Nakanishi, Sugiyama, and Kitano}}]{kobayashi2010direct}
\bibinfo{author}{\bibfnamefont{H.}~\bibnamefont{Kobayashi}},
  \bibinfo{author}{\bibfnamefont{S.}~\bibnamefont{Tamate}},
  \bibinfo{author}{\bibfnamefont{T.}~\bibnamefont{Nakanishi}},
  \bibinfo{author}{\bibfnamefont{K.}~\bibnamefont{Sugiyama}}, \bibnamefont{and}
  \bibinfo{author}{\bibfnamefont{M.}~\bibnamefont{Kitano}},
  \bibinfo{journal}{Phys. Rev. A} \textbf{\bibinfo{volume}{81}},
  \bibinfo{pages}{012104} (\bibinfo{year}{2010}).

\bibitem[{\citenamefont{Jozsa and Schlienz}(2000)}]{PhysRevA.62.012301}
\bibinfo{author}{\bibfnamefont{R.}~\bibnamefont{Jozsa}} \bibnamefont{and}
  \bibinfo{author}{\bibfnamefont{J.}~\bibnamefont{Schlienz}},
  \bibinfo{journal}{Phys. Rev. A} \textbf{\bibinfo{volume}{62}},
  \bibinfo{pages}{012301} (\bibinfo{year}{2000}).

\bibitem[{\citenamefont{Bergou et~al.}(2012)\citenamefont{Bergou, Futschik, and
  Feldman}}]{PhysRevLett.108.250502}
\bibinfo{author}{\bibfnamefont{J.~A.} \bibnamefont{Bergou}},
  \bibinfo{author}{\bibfnamefont{U.}~\bibnamefont{Futschik}}, \bibnamefont{and}
  \bibinfo{author}{\bibfnamefont{E.}~\bibnamefont{Feldman}},
  \bibinfo{journal}{Phys. Rev. Lett.} \textbf{\bibinfo{volume}{108}},
  \bibinfo{pages}{250502} (\bibinfo{year}{2012}).

\bibitem[{\citenamefont{Sugimoto et~al.}(2010)\citenamefont{Sugimoto,
  Hashimoto, Horibe, and Hayashi}}]{PhysRevA.82.032338}
\bibinfo{author}{\bibfnamefont{H.}~\bibnamefont{Sugimoto}},
  \bibinfo{author}{\bibfnamefont{T.}~\bibnamefont{Hashimoto}},
  \bibinfo{author}{\bibfnamefont{M.}~\bibnamefont{Horibe}}, \bibnamefont{and}
  \bibinfo{author}{\bibfnamefont{A.}~\bibnamefont{Hayashi}},
  \bibinfo{journal}{Phys. Rev. A} \textbf{\bibinfo{volume}{82}},
  \bibinfo{pages}{032338} (\bibinfo{year}{2010}).

\bibitem[{\citenamefont{Scully and Dr\"uhl}(1982)}]{PhysRevA.25.2208}
\bibinfo{author}{\bibfnamefont{M.~O.} \bibnamefont{Scully}} \bibnamefont{and}
  \bibinfo{author}{\bibfnamefont{K.}~\bibnamefont{Dr\"uhl}},
  \bibinfo{journal}{Phys. Rev. A} \textbf{\bibinfo{volume}{25}},
  \bibinfo{pages}{2208} (\bibinfo{year}{1982}).

\bibitem[{\citenamefont{Aharonov et~al.}(1988)\citenamefont{Aharonov, Albert,
  and Vaidman}}]{PhysRevLett.60.1351}
\bibinfo{author}{\bibfnamefont{Y.}~\bibnamefont{Aharonov}},
  \bibinfo{author}{\bibfnamefont{D.~Z.} \bibnamefont{Albert}},
  \bibnamefont{and} \bibinfo{author}{\bibfnamefont{L.}~\bibnamefont{Vaidman}},
  \bibinfo{journal}{Phys. Rev. Lett.} \textbf{\bibinfo{volume}{60}},
  \bibinfo{pages}{1351} (\bibinfo{year}{1988}).

\bibitem[{\citenamefont{Kobayashi
  et~al.}(2011{\natexlab{a}})\citenamefont{Kobayashi, Tamate, Nakanishi,
  Sugiyama, and Kitano}}]{JPSJ.80.034401}
\bibinfo{author}{\bibfnamefont{H.}~\bibnamefont{Kobayashi}},
  \bibinfo{author}{\bibfnamefont{S.}~\bibnamefont{Tamate}},
  \bibinfo{author}{\bibfnamefont{T.}~\bibnamefont{Nakanishi}},
  \bibinfo{author}{\bibfnamefont{K.}~\bibnamefont{Sugiyama}}, \bibnamefont{and}
  \bibinfo{author}{\bibfnamefont{M.}~\bibnamefont{Kitano}},
  \bibinfo{journal}{J. Phys. Soc. Jpn.} \textbf{\bibinfo{volume}{80}},
  \bibinfo{pages}{034401} (\bibinfo{year}{2011}{\natexlab{a}}).

\bibitem[{\citenamefont{Tamate et~al.}(2009)\citenamefont{Tamate, Kobayashi,
  Nakanishi, Sugiyama, and Kitano}}]{tamate2009geometrical}
\bibinfo{author}{\bibfnamefont{S.}~\bibnamefont{Tamate}},
  \bibinfo{author}{\bibfnamefont{H.}~\bibnamefont{Kobayashi}},
  \bibinfo{author}{\bibfnamefont{T.}~\bibnamefont{Nakanishi}},
  \bibinfo{author}{\bibfnamefont{K.}~\bibnamefont{Sugiyama}}, \bibnamefont{and}
  \bibinfo{author}{\bibfnamefont{M.}~\bibnamefont{Kitano}},
  \bibinfo{journal}{New J. Phys.} \textbf{\bibinfo{volume}{11}},
  \bibinfo{pages}{093025} (\bibinfo{year}{2009}).

\bibitem[{\citenamefont{Schmitzer et~al.}(1993)\citenamefont{Schmitzer, Klein,
  and Dultz}}]{PhysRevLett.71.1530}
\bibinfo{author}{\bibfnamefont{H.}~\bibnamefont{Schmitzer}},
  \bibinfo{author}{\bibfnamefont{S.}~\bibnamefont{Klein}}, \bibnamefont{and}
  \bibinfo{author}{\bibfnamefont{W.}~\bibnamefont{Dultz}},
  \bibinfo{journal}{Phys. Rev. Lett.} \textbf{\bibinfo{volume}{71}},
  \bibinfo{pages}{1530} (\bibinfo{year}{1993}).

\bibitem[{\citenamefont{Tewari et~al.}(1995)\citenamefont{Tewari, Ashoka, and
  Sree~Ramana}}]{tewari1995four}
\bibinfo{author}{\bibfnamefont{S.~P.} \bibnamefont{Tewari}},
  \bibinfo{author}{\bibfnamefont{V.}~\bibnamefont{Ashoka}}, \bibnamefont{and}
  \bibinfo{author}{\bibfnamefont{M.}~\bibnamefont{Sree~Ramana}},
  \bibinfo{journal}{Opt. Commun.} \textbf{\bibinfo{volume}{120}},
  \bibinfo{pages}{235} (\bibinfo{year}{1995}).

\bibitem[{\citenamefont{Li et~al.}(1999)\citenamefont{Li, Gong, Gao, and
  Chen}}]{li1999experimental}
\bibinfo{author}{\bibfnamefont{Q.}~\bibnamefont{Li}},
  \bibinfo{author}{\bibfnamefont{L.}~\bibnamefont{Gong}},
  \bibinfo{author}{\bibfnamefont{Y.}~\bibnamefont{Gao}}, \bibnamefont{and}
  \bibinfo{author}{\bibfnamefont{Y.}~\bibnamefont{Chen}},
  \bibinfo{journal}{Opt. Commun.} \textbf{\bibinfo{volume}{169}},
  \bibinfo{pages}{17} (\bibinfo{year}{1999}).

\bibitem[{\citenamefont{Hils et~al.}(1999)\citenamefont{Hils, Dultz, and
  Martienssen}}]{PhysRevE.60.2322}
\bibinfo{author}{\bibfnamefont{B.}~\bibnamefont{Hils}},
  \bibinfo{author}{\bibfnamefont{W.}~\bibnamefont{Dultz}}, \bibnamefont{and}
  \bibinfo{author}{\bibfnamefont{W.}~\bibnamefont{Martienssen}},
  \bibinfo{journal}{Phys. Rev. E} \textbf{\bibinfo{volume}{60}},
  \bibinfo{pages}{2322} (\bibinfo{year}{1999}).

\bibitem[{\citenamefont{Tamate et~al.}(2011)\citenamefont{Tamate, Ogawa, and
  Kitano}}]{tamate2011bloch}
\bibinfo{author}{\bibfnamefont{S.}~\bibnamefont{Tamate}},
  \bibinfo{author}{\bibfnamefont{K.}~\bibnamefont{Ogawa}}, \bibnamefont{and}
  \bibinfo{author}{\bibfnamefont{M.}~\bibnamefont{Kitano}},
  \bibinfo{journal}{Phys. Rev. A} \textbf{\bibinfo{volume}{84}},
  \bibinfo{pages}{052114} (\bibinfo{year}{2011}).

\bibitem[{\citenamefont{Bogdanov et~al.}(2004)\citenamefont{Bogdanov, Chekhova,
  Kulik, Maslennikov, Zhukov, Oh, and Tey}}]{Physrevlett.93.230503}
\bibinfo{author}{\bibfnamefont{Y.~I.} \bibnamefont{Bogdanov}},
  \bibinfo{author}{\bibfnamefont{M.~V.} \bibnamefont{Chekhova}},
  \bibinfo{author}{\bibfnamefont{S.~P.} \bibnamefont{Kulik}},
  \bibinfo{author}{\bibfnamefont{G.~A.} \bibnamefont{Maslennikov}},
  \bibinfo{author}{\bibfnamefont{A.~A.} \bibnamefont{Zhukov}},
  \bibinfo{author}{\bibfnamefont{C.~H.} \bibnamefont{Oh}}, \bibnamefont{and}
  \bibinfo{author}{\bibfnamefont{M.~K.} \bibnamefont{Tey}},
  \bibinfo{journal}{Phys. Rev. Lett.} \textbf{\bibinfo{volume}{93}},
  \bibinfo{pages}{230503} (\bibinfo{year}{2004}).

\bibitem[{\citenamefont{Lanyon et~al.}(2008)\citenamefont{Lanyon, Weinhold,
  Langford, O'Brien, Resch, Gilchrist, and White}}]{PhysRevLett.100.060504}
\bibinfo{author}{\bibfnamefont{B.~P.}~\bibnamefont{Lanyon}},
  \bibinfo{author}{\bibfnamefont{T.~J.}~\bibnamefont{Weinhold}},
  \bibinfo{author}{\bibfnamefont{N.~K.}~\bibnamefont{Langford}},
  \bibinfo{author}{\bibfnamefont{J.~L.}~\bibnamefont{O'Brien}},
  \bibinfo{author}{\bibfnamefont{K.~J.}~\bibnamefont{Resch}},
  \bibinfo{author}{\bibfnamefont{A.}~\bibnamefont{Gilchrist}},
  \bibnamefont{and} \bibinfo{author}{\bibfnamefont{A.~G.}~\bibnamefont{White}},
  \bibinfo{journal}{Phys. Rev. Lett.} \textbf{\bibinfo{volume}{100}},
  \bibinfo{pages}{060504} (\bibinfo{year}{2008}).

\bibitem[{\citenamefont{D'Ariano et~al.}(2005)\citenamefont{D'Ariano, Mataloni,
  and Sacchi}}]{PhysRevA.71.062337}
\bibinfo{author}{\bibfnamefont{G.~M.}~\bibnamefont{D'Ariano}},
  \bibinfo{author}{\bibfnamefont{P.}~\bibnamefont{Mataloni}}, \bibnamefont{and}
  \bibinfo{author}{\bibfnamefont{M.~F.}~\bibnamefont{Sacchi}},
  \bibinfo{journal}{Phys. Rev. A} \textbf{\bibinfo{volume}{71}},
  \bibinfo{pages}{062337} (\bibinfo{year}{2005}).

\bibitem[{\citenamefont{Vallone et~al.}(2007)\citenamefont{Vallone, Pomarico,
  De~Martini, Mataloni, and Barbieri}}]{PhysRevA.76.012319}
\bibinfo{author}{\bibfnamefont{G.}~\bibnamefont{Vallone}},
  \bibinfo{author}{\bibfnamefont{E.}~\bibnamefont{Pomarico}},
  \bibinfo{author}{\bibfnamefont{F.}~\bibnamefont{De~Martini}},
  \bibinfo{author}{\bibfnamefont{P.}~\bibnamefont{Mataloni}}, \bibnamefont{and}
  \bibinfo{author}{\bibfnamefont{M.}~\bibnamefont{Barbieri}},
  \bibinfo{journal}{Phys. Rev. A} \textbf{\bibinfo{volume}{76}},
  \bibinfo{pages}{012319} (\bibinfo{year}{2007}).

\bibitem[{\citenamefont{Ogawa et~al.}(2013)\citenamefont{Ogawa, Tamate,
  Kobayashi, Nakanishi, and Kitano}}]{PhysRevA.88.063813}
\bibinfo{author}{\bibfnamefont{K.}~\bibnamefont{Ogawa}},
  \bibinfo{author}{\bibfnamefont{S.}~\bibnamefont{Tamate}},
  \bibinfo{author}{\bibfnamefont{H.}~\bibnamefont{Kobayashi}},
  \bibinfo{author}{\bibfnamefont{T.}~\bibnamefont{Nakanishi}},
  \bibnamefont{and} \bibinfo{author}{\bibfnamefont{M.}~\bibnamefont{Kitano}},
  \bibinfo{journal}{Phys. Rev. A} \textbf{\bibinfo{volume}{88}},
  \bibinfo{pages}{063813} (\bibinfo{year}{2013}).

\bibitem[{\citenamefont{Kobayashi
  et~al.}(2011{\natexlab{b}})\citenamefont{Kobayashi, Ikeda, Tamate, Nakanishi,
  and Kitano}}]{kobayashi2011nonlinear}
\bibinfo{author}{\bibfnamefont{H.}~\bibnamefont{Kobayashi}},
  \bibinfo{author}{\bibfnamefont{Y.}~\bibnamefont{Ikeda}},
  \bibinfo{author}{\bibfnamefont{S.}~\bibnamefont{Tamate}},
  \bibinfo{author}{\bibfnamefont{T.}~\bibnamefont{Nakanishi}},
  \bibnamefont{and} \bibinfo{author}{\bibfnamefont{M.}~\bibnamefont{Kitano}},
  \bibinfo{journal}{Phys. Rev. A} \textbf{\bibinfo{volume}{83}},
  \bibinfo{pages}{063808} (\bibinfo{year}{2011}{\natexlab{b}}).

\bibitem[{\citenamefont{Brendel et~al.}(1995)\citenamefont{Brendel, Dultz, and
  Martienssen}}]{PhysRevA.52.2551}
\bibinfo{author}{\bibfnamefont{J.}~\bibnamefont{Brendel}},
  \bibinfo{author}{\bibfnamefont{W.}~\bibnamefont{Dultz}}, \bibnamefont{and}
  \bibinfo{author}{\bibfnamefont{W.}~\bibnamefont{Martienssen}},
  \bibinfo{journal}{Phys. Rev. A} \textbf{\bibinfo{volume}{52}},
  \bibinfo{pages}{2551} (\bibinfo{year}{1995}).

\bibitem[{\citenamefont{Loredo et~al.}(2014)\citenamefont{Loredo, Broome,
  Smith, and White}}]{PhysRevLett.112.143603}
\bibinfo{author}{\bibfnamefont{J.~C.} \bibnamefont{Loredo}},
  \bibinfo{author}{\bibfnamefont{M.~A.} \bibnamefont{Broome}},
  \bibinfo{author}{\bibfnamefont{D.~H.} \bibnamefont{Smith}}, \bibnamefont{and}
  \bibinfo{author}{\bibfnamefont{A.~G.} \bibnamefont{White}},
  \bibinfo{journal}{Phys. Rev. Lett.} \textbf{\bibinfo{volume}{112}},
  \bibinfo{pages}{143603} (\bibinfo{year}{2014}).

\bibitem[{\citenamefont{Majorana}(1932)}]{majorana}
\bibinfo{author}{\bibfnamefont{E.}~\bibnamefont{Majorana}},
  \bibinfo{journal}{Nuovo Cimento} \textbf{\bibinfo{volume}{9}},
  \bibinfo{pages}{43} (\bibinfo{year}{1932}).

\bibitem[{\citenamefont{Liu and Fu}(2014)}]{PhysRevLett.113.240403}
\bibinfo{author}{\bibfnamefont{H.~D.} \bibnamefont{Liu}} \bibnamefont{and}
  \bibinfo{author}{\bibfnamefont{L.~B.} \bibnamefont{Fu}},
  \bibinfo{journal}{Phys. Rev. Lett.} \textbf{\bibinfo{volume}{113}},
  \bibinfo{pages}{240403} (\bibinfo{year}{2014}).

\bibitem[{\citenamefont{Bruno}(2012)}]{PhysRevLett.108.240402}
\bibinfo{author}{\bibfnamefont{P.}~\bibnamefont{Bruno}},
  \bibinfo{journal}{Phys. Rev. Lett.} \textbf{\bibinfo{volume}{108}},
  \bibinfo{pages}{240402} (\bibinfo{year}{2012}).

\bibitem[{\citenamefont{Hannay}(1998)}]{hannay1998berry}
\bibinfo{author}{\bibfnamefont{J.}~\bibnamefont{Hannay}}, \bibinfo{journal}{J.
  Phys. A} \textbf{\bibinfo{volume}{31}}, \bibinfo{pages}{L53}
  (\bibinfo{year}{1998}).

\end{thebibliography}

\end{document}